\documentclass[a4paper]{jpconf}
\usepackage{graphicx}
\usepackage{bm}
\usepackage{amsmath}
\usepackage{amssymb}

\begin{document}

\title{Superconducting pairing of interacting electrons: implications from the two-impurity Anderson model}
\author{Lijun Zhu and Jian-Xin Zhu}
\address{Theoretical Division and Center for Nonlinear Studies, Los Alamos National Laboratory,
Los Alamos, New Mexico 87545, USA}
\ead{ljzhu@lanl.gov}
\begin{abstract}

We study the non-local superconducting pairing of two interacting Anderson impurities, which has an instability near the quantum critical point from the competition between the Kondo effect and an antiferromagnetic inter-impurity spin exchange interaction.  As revealed by the dynamics over the whole energy range, the superconducting pairing fluctuations acquire considerable strength from an energy scale much higher than the characteristic spin fluctuation scale while the low energy behaviors follow those of the staggered spin susceptibility. We argue that the glue to the superconducting pairing is not the spin fluctuations, but rather the effective Coulomb interaction. On the other hand, critical spin fluctuations in the vicinity of quantum criticality are also crucial to a superconducting pairing instability, by preventing a Fermi liquid fixed point being reached to keep the superconducting pairing fluctuations finite at low energies. A superconducting order, to reduce the accumulated entropy carried by the critical degrees of freedom, may arise favorably from this instability.

\end{abstract}

\section{Introduction}
\label{sec:intro}
In conventional superconductors, the electron pairing arises from the mediation of lattice vibrations~\cite{RDParks}. In cuprate superconductors, the pairing mechanism for superconductivity has been hotly debated as to whether the pairing occurs via the coupling between fermionic quasiparticles and bosonic modes, or whether a bosonic glue is really necessary~\cite{PWAnderson}.  In the former type of mechanism, the scenario of electronic coupling to spin fluctuations is the leading contender~\cite{TAMaier}. However, in high-$T_c$ cuprates, the antiferromagnetic N\'{e}el state is terminated far before the unconventional superconductivity emerges as the hole doping varies. As such, the relation between a magnetic quantum criticality point and high-$T_c$ superconductivity has not been developed. Furthermore, the final identification of the mechanism is also complicated by the non-negligible electron-phonon coupling as revealed in various experiments~\cite{TCuk}.
In this aspect, heavy fermion materials have been providing test grounds for novel mechanism for superconductivity~\cite{Stewart01}. In these systems, the interplay between magnetism and superconductivity has been observed unambiguously. The generic phase diagram shows that, when a tuning parameter such as doping, magnetic field, and pressure is varied, the magnetic state is either interrupted by or in close vicinity to the appearance of a superconducting state. Therefore, they are the most ideal systems for the study on unconventional superconductivity associated with strong electron correlations in general and the quantum criticality in specific.

To address the relation between superconductivity and magnetic quantum criticality, we consider here a two-impurity Anderson model, which is a minimal setting for antiferromagnetism and non-local superconducting pairing. It is also the quantum impurity/cluster model for the periodical Anderson lattice model or the Hubbard model within the cluster dynamical mean-field theoretical approach~\cite{Kotliar05}. This model captures the essence of the competition between the local Kondo coupling between spins of the localized electrons to a conduction band, and the non-local spin exchange interaction between spins of localized electrons, e.g., Ruderman-Kittel-Kasuya-Yosida (RKKY) interaction in heavy fermion materials. Due to this competition, a quantum phase transition occurs between a Kondo resonance state and an inter-impurity spin singlet state~\cite{Jones87,Sakai89,Jones89,Affleck92,Sire93,Fye94,Gan95,Silva96}, and multiple energy scales can show up~\cite{ZhuSq10}. In this paper, we discuss the superconducting pairing between two Anderson orbitals and its mechanism.

\section{Model}
\label{sec:model}

The Hamiltonian for the two-impurity Anderson model can be written as
\begin{eqnarray}
H &=&   \sum_{{\bf k}\sigma} \epsilon_{\bf k} c^\dag_{{\bf k}\sigma} c_{{\bf k}\sigma} + \sum_{{\bf k}\sigma i} (V_{\bf k}e^{i {\bf k}\cdot {\bf r}_i}
c^\dag_{{\bf k}\sigma}f_{i\sigma} + \text{h.c.})  
+\sum_{i\sigma} (\epsilon_f  f^\dag_{i\sigma} f_{i\sigma} +  {U\over 2} n_{f i\sigma}n_{f i{\bar \sigma}}) + I {\bf S}_1\cdot {\bf S}_2\;,
\label{eq:hamiltonian}
\end{eqnarray}
which describes two interacting local orbitals  $f_{i\sigma}$ ($i=1,2$), with energy level $\epsilon_f$ and onsite Coulomb repulsion $U$, in hybridization with a non-interacting conduction electron band $c_{{\bf k}\sigma}$ with the strength $V_{\bf k}$ at each impurity site ${\bf r}_i$.  In terms of the even or odd parity combinations $f_{p=(e,o),\sigma} = (f_{1\sigma}\pm f_{2\sigma})/\sqrt{2}$, the fluctuations due to conduction electrons can be represented by two separate baths with the hybridization functions $\Gamma_{e,o}(\omega)   =  -{1\over 2} \text{Im} [ \sum_{\bf k} V^2_{\bf k} |e^{i{\bf k}\cdot {\bf r}_1} \pm e^{i{\bf k}\cdot {\bf r}_2}|^2 / (\omega -\epsilon_{\bf k} + i 0^+)] $. In a realistic system RKKY interaction is perturbatively generated. Here we consider a case $\Gamma_{e,o}(\omega) = \Gamma_0$ for $|\omega| <D$ ($\Gamma_0$ is a constant and $D$ the half bandwidth of the conduction electrons), for which the generated RKKY interaction vanishes,  and we tune the intersite spin interaction by a direct RKKY term $ I {\bf S}_1\cdot {\bf S}_2$.

We focus on a non-local superconducting pairing correlation function $\chi_{sc}=\langle\langle \hat{\Delta}^{\dagger}; \hat{\Delta} \rangle\rangle$ with $\hat{\Delta}^{\dagger}=(f^\dag_{1\uparrow}f^\dag_{2\downarrow} + f^\dag_{2\uparrow} f^\dag_{1\downarrow})/\sqrt{2}$, i.e., in a singlet configuration. To illustrate its relation to spin and charge fluctuations, we also provide results on the uniform and staggered charge susceptibilities $\chi_{c(u,a)} = \langle\langle n_{1}\pm  n_2; n_{1}\pm n_2\rangle\rangle/2$, the uniform and staggered spin susceptibilities $\chi_{u,a} = \langle\langle S_{1z}\pm  S_{2z}; S_{1z}\pm S_{2z}\rangle\rangle/2$, the current fluctuation $\chi_{curr}$ with $\hat{\Delta}_{curr}^{\dagger}=i\sum_{\sigma}(f^\dag_{1\sigma}f_{2\sigma} - f^\dag_{2\sigma} f_{1\sigma})/\sqrt{2}$, as well as the single-particle properties including the spectral function $A_f= -(1/\pi) \text{Im}G_f$ and the interaction part of the self energy $-\text{Im}\Sigma_f$. These quantities are calculated from the complete-Fock-space numerical renormalization group method (CFS-NRG)~\cite{Anders06} while a two-particle Green's function method~\cite{Bulla} is adopted for calculating the self energy.

\section{Correlation fluctuations and the associated energy scales}
\label{sec:twoscales}

\begin{figure}[tbh]
\centering
\includegraphics[width=\columnwidth]{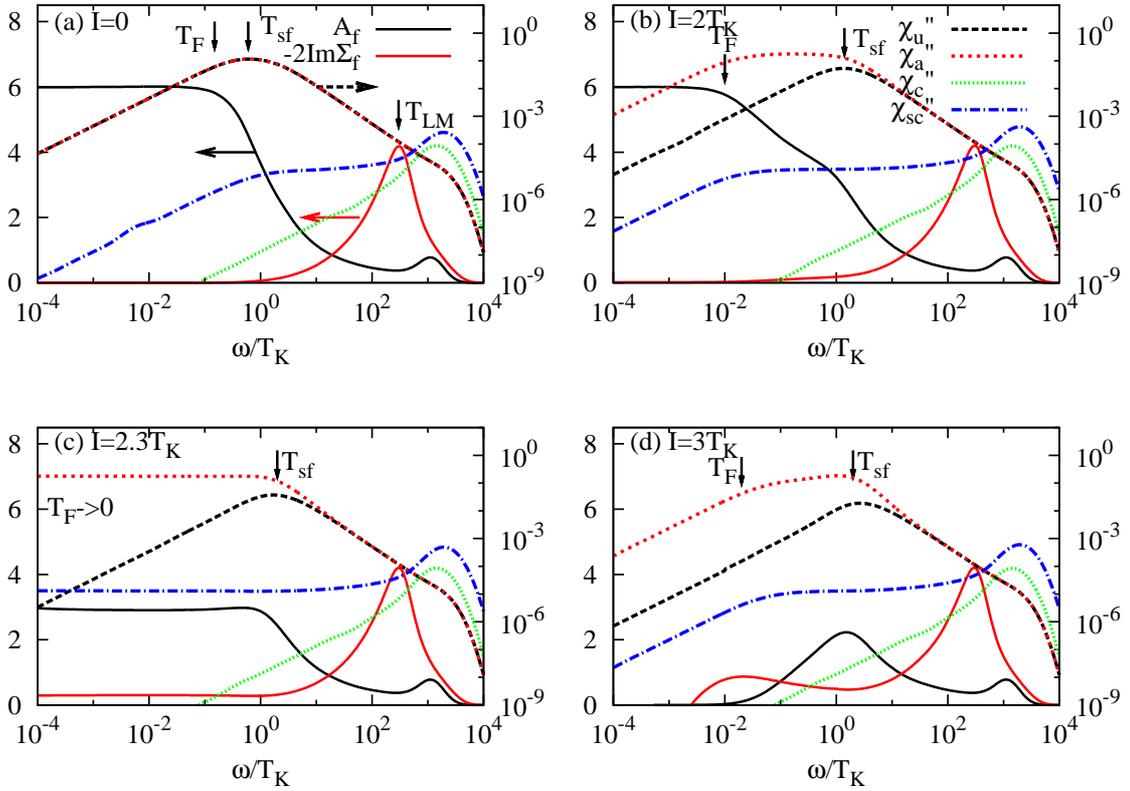}
\caption{ Dynamical properties of the two-impurity Anderson model for various values of the inter-impurity spin exchange interaction $I$: (a) $I=0$, (b) $I<I_c$, (c) $I\approx I_c$ and (d) $I>I_c$. The spectral functions $A_f$ (in unit of $D$), and the imaginary part of the self-energy $-\text{Im}\Sigma_f$ (in unit of $1/D$) are shown in log-linear scale (left y-axis), while the imaginary parts of the uniform spin susceptibility $\chi''_u$, the staggered spin susceptibility $\chi''_a$, the uniform and staggered charge susceptibilities (the same in this case) $\chi''_c$, and the superconducting pairing fluctuation $\chi''_{sc}$ (also the half of the current fluctuation $\chi_{curr}$ due to the axial charge symmetry), all in unit of $1/T_K$, are shown in log-log scale (right y-axis).  Different energy scales, such as the Fermi liquid scale $T_F$, the spin-fluctuation scale $T_{sf}$, and the local moment scale $T_{LM}$ are labeled.  Here $\Gamma_{e,o} =0.045\pi D$ and $\epsilon_f = -U/2=-D \approx -10^3 T_K$. In this case, $A_f$ and $-\text{Im}\Sigma_f$ are found to be the same for different parities and spins.
}
\label{fig:corrfunc}
\end{figure}

The results for these dynamical quantities for different values of $I$ are shown in Fig.~\ref{fig:corrfunc}. Some of the calculation details can be found in Ref.~\cite{ZhuSq10}. With the tuning of an antiferromagnetic $I$, there is a change of the ground state from the Kondo resonance state $A_f(\omega=0) \approx 1/(\pi \Gamma_0)$ [e.g., Fig.~\ref{fig:corrfunc}(a-b)], to the inter-impurity spin singlet state with a pseudogap $A_f(\omega)\sim \omega^2$ [e.g., Fig.~\ref{fig:corrfunc}(d)]. Though the spectral weight has a sudden change at the Fermi energy, this transition is continuous as the effective (local) Fermi liquid (FL) temperature $T_F$ (both competing states belong to FL fixed points) vanishes uniformly at the critical point $I_c \approx 2.3 T_K$ [Fig.~\ref{fig:corrfunc}(c)], where $T_K$ is the single-impurity Kondo temperature. $\chi_u$ and $\chi_a$ have different behaviors.  While $ \chi_u''(\omega)$ can be characterized by one monotonically increasing energy scale $T_{sf}$, $\chi_a''(\omega)$ in general has a flattened peak spanning a finite energy range from $T_F$ to $T_{sf}$. ($T_F$ and $T_{sf}$ correspond to $T_L$ and $T_H$ in Ref.~\cite{ZhuSq10}). $T_{sf}$ therefore characterizes a spin fluctuation scale, which takes the value between $T_K$ and $I$. Between $T_F$ and $T_{sf}$, there are still finite spectral weights but these quasiparticles develop non-Fermi liquid behaviors, which relate to partially screened Kondo resonances~\cite{ZhuSq10}.

We now examine other correlation functions. For the charge fluctuations, we find that they play little role in the low energy physics. They have resonance peaks around $U/2$ and decrease as $\sim\omega$ at low energies from a scale $T_{LM}$. Here, the uniform and staggered charge susceptibilities are found to be the same and remain unchanged with the tuning of $I$. (However, in the particle-hole asymmetric case, it is found that the uniform charge susceptibility also diverges at the quantum critical point~\cite{ZhuVarmaZhu,Hattori}). For the superconducting (SC) pairing correlation function $\chi_{sc}(\omega)$, we find the low energy part of $\chi''_{sc}(\omega)$ follows the behavior of $\chi''_a(\omega)$, as $\sim\omega$ for $\omega < T_F$ and as $\sim C_{sc}$ for $\omega > T_F$, where $C_{sc}$ is relatively a constant. At the critical point, as $T_F\to 0$, $\chi'_{sc}(0)$ becomes divergent, leading to a SC pairing instability. However, the high energy part doesn't follow the spin fluctuations: it has a resonance peak at $U$ and the SC fluctuation softening happens from $T_{LM}$, which is much larger than $T_{sf}$, down to $T_F$, or $\chi'_{sc}(0) \approx C_{sc} \ln (T_{LM}/T_F)$.

$T_{LM}$ is the also the energy where $-\text{Im}\Sigma(\omega)$ reaches a maximal value. In the single-impurity Anderson model~\cite{HRKrishna-murthy},  the renormalization group (RG) flow starts from the free orbital fixed point, rather than directly from the local moment fixed point as in the Kondo model. The free orbital fixed point is unstable against any interaction $U$, or the hybridization with the conduction electrons $\Gamma$. $U_{eff}$ and $\Gamma_{eff}$ increase at first, approaching the local moment fixed point, which is characterized by $U_{eff} \to \infty$ and a small Kondo coupling $ J_{K,eff} = \Gamma_{eff}/U_{eff}$, and then $U_{eff}$ decreases while $\Gamma_{eff}$ keeps increasing, reaching the strong coupling fixed point $J_{K,eff} \to \infty$ (due to the marginal nature of  $J_K$, such change is slow and only becomes abrupt near $T_K$).  $T_{LM}$ therefore characterizes the proximity to the local moment fixed point in the RG flow.

\section{On the superconducting pairing}
\label{sec:sc}
\begin{figure}[tbh]
\begin{center}
    \begin{minipage}{0.5\textwidth}
    \includegraphics[width=\columnwidth]{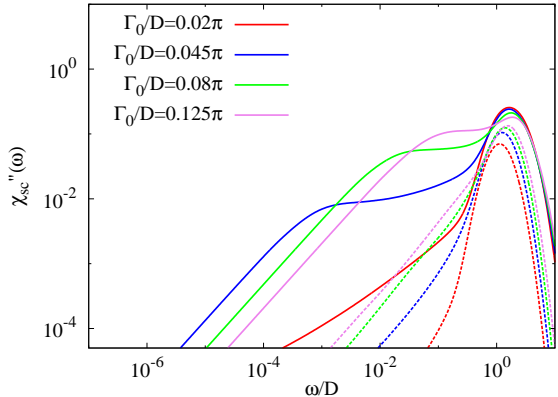}
    \end{minipage}\hfill
    \begin{minipage}{0.4\textwidth}
     \caption{ The non-local singlet ($\chi_{sc}$, solid lines) and the local ($\chi_{loc}^{sc}$, dashed lines) superconducting pairing fluctuations for different values of $\Gamma_{e,o}(\omega)=\Gamma_0$. $U$ is fixed as $2D$, as in Fig. 1, and $I=0$.}
\label{fig:scfunc}
    \end{minipage}
  \end{center}
\end{figure}

The fact that the SC pairing fluctuations become significant before either $T_K$ or $I$,  and in addition their weight remains relatively unchanged at $\omega =T_{sf}$ for finite $I$s, implies that the SC pairing might not require the spin fluctuations as the glue. The central issue is therefore to understand the SC pairing fluctuation softening beginning at $T_{LM}$. In terms of the even and odd parity orbitals, the interaction part of the original Hamiltonian can be rewritten as
\begin{eqnarray}
H_{f,int} &=&  ( {U/2}- {3I /8}) ( n_{e\uparrow} n_{e\downarrow} + n_{o\uparrow} n_{o\downarrow})
+(  {U/2}+{3I/8} ) {\bf J}_{e} \cdot {\bf J}_o  - (U/2-I/8) {\bf S}_{e} \cdot {\bf S}_o,
\label{eq:hameo}
\end{eqnarray}
where ${\bf J}$ and ${\bf S}$ are axial charge~\cite{Jones87} and spin of the even and odd orbitals. For $J^+_p = f^\dag_{p\uparrow}f^\dag_{p\downarrow}$, it can be easily shown that the transverse ``staggered" and ``uniform" axial charge susceptibilities, $\langle\langle J^+_e\mp J^+_o; J^-_e\mp J^-_o\rangle\rangle$ correspond to the non-local singlet SC pairing correlation function and the local SC pairing correlation function $\chi_{loc}^{sc} =\langle\langle \sum_i f^\dag_{i\uparrow}f^\dag_{i\downarrow}; \sum_i f_{i\downarrow}f_{i\uparrow}\rangle\rangle$. Due to the axial charge rotational symmetry, they are the same as twice of the correlation functions of $J_{e}^z \mp J_{o}^z=(n_{e}\mp n_o)/2$, which relate to the current fluctuations and uniform charge susceptibility, respectively (explicitly verified by calculations). With an ``antiferromagnetic" axial charge exchange interaction $I_{ac} = U+3I/4$, the staggered component has negative energy while the uniform component has positive energy, i.e, the non-local SC pairing is energetically favored against the local SC pairing. Such axial charge fluctuations are also subjected to the coupling to the conduction electron bath. In Fig.~\ref{fig:scfunc}, we show the behaviors of $\chi''_{sc}$ and $(\chi_{loc}^{sc})''$ for different values of the hybridization constant $\Gamma_0$ ($U$ is fixed). We observe that the non-local SC pairing fluctuations are always stronger than the local ones, indicating the ``antiferromagnetic" nature. Besides, the enhancement (or the softening) of $\chi''_{sc}$ is only achieved by a relative large value of $\Gamma_0$. While a detailed analysis of this behavior is desired, we can associate it with an ``axial charge Kondo'' effect, by making an analog between the axial charge and the spin. We expect a ``Kondo"-like exchange interaction between the axial charges of the local orbitals and those of the conduction electrons in the form $J_{ac} {\bf J}_p \cdot {\bf J}_{c,p}$, and as expected, $J_{ac} \sim J_K$. Such an interaction competes with $I_{ac}$ in the same fashion as the competition between the Kondo coupling and RKKY interaction in the spin channel. Indeed, we find the behaviors of $\chi''_{sc}$ in Fig.~\ref{fig:scfunc} share similarities to the spin susceptibilities when the ratio between $T_K$ and $I$ is varied. In the spin dynamics, the softening of $\chi''_a$ is due to the equivalent strength of local Kondo splin-flip scattering and inter-impurity spin singlet to triplet scattering. The same type of physics is expected for the axial charge fluctuations: the axial charge fluctuations with conduction electrons compete with the axial charge fluctuations between non-local (``singlet'') and local (``triplet'') SC pairs.

On the other hand, the low energy edge of the SC pairing fluctuation softening is $T_F$, rather than $T_{sf}$ for finite $I$s. This indicates that the SC pairing fluctuations are only suppressed when the strong coupling fixed point, or a FL fixed point is reached. This can be understood as that the effective Kondo couplings in the spin and axial charge channels are the same.  When the strong coupling fixed point is reached, $J_{K,eff}\to \infty$, the axial charge for each parity forms an ``axial charge Kondo singlet'' with those of the conduction electrons. The axial charge fluctuations between even and odd parities, i.e., the SC pairing fluctuations are suppressed. While the spin exchange interaction $I$ (antiferromagnetic in our study, but we expect similar physics for the triplet pairing with ferromagnetic $I$) can create an ``antiferromagnetic'' interaction ($I_{ac}$) in the axial charge channel as well [Cf. Eq.(\ref{eq:hameo})], but in general, $I \ll U$. This also explains that $\chi''_{sc}(\omega)$ has little change at $\omega = I$. The effect of $I$ in this picture is rather to compete with the Kondo effect and keep $J_{K,eff}$ finite at low energies. This implies that the SC pairing can only keep its strength to low energies when the Fermi liquid fixed point is avoided, i.e., near the quantum critical point. This can also be evidenced from a low energy effective Hamiltonian in terms of Fermi liquid parameters~\cite{Jones87}, where $I_{ac, eff}$ is found to be divergent at the quantum critical point.

In the traditional theory of superconductivity, the formation of Cooper pairs requires an attractive interaction, direct or generated with certain glues. Our results rather imply that such an effective attractive interaction in the non-local singlet channel is originated from the onsite repulsive interaction. The SC pairing fluctuations at high energies may simply be two-particle excitation fluctuations.  The ``antiferromagnetic'' nature of $I_{ac}$, which appears as an attractive interaction in the site basis $-(I_{ac}/2)(n_{1\uparrow}n_{2\downarrow}+n_{2\uparrow}n_{1\downarrow})$, may be due to the repulsive nature of local Coulomb interaction, i.e., the non-local two-particle excitations are favored against local pairs. The central issue revealed by this study is that such two-particle excitations can be preserved to low energies. This is achieved by an ``axial charge Kondo effect'' to enhance pair fluctuations with coupling to conduction electrons. Such an enhancement maintains further to the Fermi energy when the spin dynamics are critical near a quantum critical point. We  notice that the spin fluctuations are the dominating fluctuations at low energies. In other words, the weights of the low energy states are dominated by the single-occupancy states (in the site basis). The SC pairing fluctuations are subjected to the weight of the double or zero occupancy configurations in the low energy states, which is $\sim(\Gamma/U)^2$. Therefore, the effective SC pairing energy scale is $I_{ac, eff}=U (\Gamma/U)^2 = \Gamma^2/U$ instead of $U$. The dominating excitations would still be single-particle excitations. However, the single-particle excitations at quantum criticality have non-Fermi liquid behaviors. In the two-impurity model, they are formulated as Majorana Fermions~\cite{Affleck92,Sire93,Gan95}. The origin for Majorana Fermions is also related to the finite $J_{K,eff}$, which can only lead to partial-screened Kondo resonances. They carry finite residue entropies~\cite{Affleck92,Sire93,Gan95}.  A superconducting order, by forming Cooper pairs which is a lower entropy state, becomes favorable to these single-particle excitations. We notice that the entropy accumulation is also a generic thermodynamic feature of quantum criticality~\cite{Zhu03}.

In summary, from the dynamical property of the superconducting pairing correlation function in the two-impurity Anderson model, we learn that the superconducting pairing doesn't necessarily need spin fluctuations, or bosonic modes as the glue. It can be originated from the effective Coulomb interaction. However, such pairing strength at low energies only remains finite when spin fluctuations are critical to avoid a Fermi liquid fixed point. This is manifested as being in the vicinity of certain quantum critical points. This provides us a new picture on the origin of unconventional superconductivity in strongly correlated electron systems. It will be interesting to check how this idea can be applied to the lattice systems.

\ack
We would like to thank Qimiao Si, Joe D. Thompson and Chandra M. Varma for helpful discussions. This work was supported by the U.S. DOE  through the LDRD Program at LANL.

\end{document}